\documentclass[12pt]{article}

\usepackage{graphicx}
\usepackage{dcolumn}
\usepackage{bm}

\author{
        Edmanuel Torres \\
        National Institute for Nanotechnology, 11421 \\
        Saskatchewan Drive, Edmonton, Alberta, Canada, \\ 
        also at the Faculty of Basic Sciences, Universidad \\
        Tecnol\'ogica de Bol\'ivar, Cartagena, Colombia \\
        and \\
        Rafael Torres \\
        Department of Physics, Universidad Industrial de Santander,\\
        Bucaramanga, Colombia
}

\title{Chirp-wave Expansion of the Electron Wavefunctions in Atoms}%

\date{\today}

\begin{document}

\maketitle

\begin{abstract}
The description of the electron wavefunctions in atoms is generalized to the fractional Fourier series. This method introduces a continuous and infinite number of chirp basis sets with linear variation of the frequency to expand the wavefunctions, in which plane-waves are a special case. The chirp characteristics of each basis set can be adjusted through a single parameter. Thus, the basis set cutoff can be optimized variationally. The approach is tested with the expansion of the electron wavefunctions in atoms, and it is shown that chirp basis sets substantially improve the convergence in the description of the electron density. We have found that the natural oscillations of the electron core states are efficiently described in chirp-waves.
\end{abstract}

\section{Introduction}
The electrons in periodic potentials are described by Bloch waves~\cite{Bloch:1928}, which are in general expanded on an auxiliary  set of basis functions. Plane-waves are in particular convenient because they are periodic and conform a complete orthonormal basis set, therefore they do not suffer of basis set superposition error. However, plane-waves are inefficient in describing the radial nodes of the wavefunctions. In particular, the kinetic energy of valence electrons is substantially increased due the orthogonality of their wavefunctions inside the atomic core region. In order to minimize this inconvenience, methods to decouple the core and valence states have been developed. Slater~\cite{Slater:1937} used radial solutions inside spheres surrounding the atoms, reducing the number of plane-waves to only the necessary to describe the {valence electrons} in between the spheres. Herring further developed the concept by orthogonalizing each valence plane-wave function to all core wavefunctions~\cite{Herring:1940}, and later formalized as a pseudopotential theory by Phillips and Kleinman~\cite{Phillips:1959}.

The number of auxiliary basis function directly impact the computational cost of the electronic structure calculations, and ultimately restrict the systems that can be efficiently modeled. This has motivated the expansion of Bloch waves on alternative basis functions~\cite{Kohanoff:2006}. In particular, Gygi has recently reformulated the plane-wave approach in curvilinear coordinates in which the adaptive Riemannian metric and plane-wave cutoff are treated variational~\cite{Gygi:1992}. However, until now plane-waves in combination with pseudopotentials have been the most widely adopted approach, to reduce the computational complexity of the Schr\"odinger equation for electrons in periodic potentials. Nevertheless, basis sets that can more efficiently describe the electron wavefunctions may foster the capabilities of electronic structure calculations.

In this article, we present the formalism for the expansion of the electron wavefunctions, in atoms for the sake of brevity, using chirp-wave (ChW) basis sets. The method is based on a set of orthonormal linear chirp functions, associated to plane-waves by the fractional Fourier transform (FrFT)~\cite{Namias:1980,McBride:1987}. The resulting chirp basis functions are analogous to perform a local gauge transformation to a set of plane-wave functions. The approach is analog to Gygi generalized plane-wave method, however, in our framework the transformation is performed in the momentum coordinates.

\section{Chirp Basis Sets}
The continuous fractional Fourier transform (FrFT) \cite{Namias:1980,McBride:1987} operator $\mathcal F_\alpha$ as defined in~\cite{Ozaktas:1994} can be expressed as
\begin{equation}
\mathcal F_\alpha[f(u)]=K_\alpha \int f(u) e^{i\pi \frac{(u^2+x^2)\cos{\alpha}-2ux}{\sin\alpha}}du, \label{eq:frft}
\end{equation}
where $\alpha=a\pi/2$ is the fractional order for $0<|a|<2$. In particular for $a=1$ correspond the standard Fourier transform. The coefficient $K_{\alpha}$ is given by
\begin{equation}
K_{\alpha}=\frac{e^{i(s(\alpha)\pi/4-\alpha/2)}}{\sqrt{|\sin\alpha|}},\label{eq:coeff_frft}
\end{equation}
with $s(\alpha)=sgn(\sin{\alpha})$. Subsequently, an orthonormal set of chirp functions can be derived through the FrFT~\cite{Pei:1999}. Let $\delta_{n\tau}(u)=\delta(u-n\tau)$ be a Dirac delta distribution in the FrFT domain, with $n$ an integer number. Evaluating the $\mathcal F_\alpha$ of $\delta_{n\tau}$ we get
\begin{equation}
\mathcal F_{-\alpha}[\delta_{n\tau}](x)=K_{-\alpha} e^{-i\pi\frac{(n^2\tau^{2}+x^2)\cos\alpha-2n\tau x}{\sin\alpha}}.\label{eq:chw}
\end{equation}
For non-integer values of $a$, we obtain chirp functions with instantaneous frequency that varies linearly with the position at the chirp rate $k=\pi\cot{\alpha}$. Based on the above result, we can define a set of linear chirp functions determined by
\begin{equation}
ChW_{n\tau;\alpha}(x)=K_{-\alpha}e^{-i[\eta+(k x-G)x]},\label{eq:chw_basis_set}
\end{equation}
with constant phase $\eta=\pi n^2\tau^2\cot{\alpha}$ and starting frequency $G=(2\pi n\tau)/\sin{\alpha}$. For $\tau=1/L$ and $a=1$ the functions are plane-waves with lattice vector $L$, and $G$ the respective vectors of the reciprocal lattice. 

Using the relation $\mathcal F_{\alpha}[ChW_{n\tau,\alpha}(x)]=\delta_{n\tau}$ and the unitary property of the $\mathcal F_{\alpha}$ operator, for a non-integer $a$, we can straightforward prove their orthogonality as follows
\begin{eqnarray}
<ChW_{n\tau;\alpha}|ChW_{n'\tau;\alpha}>&=&<\mathcal F_{-\alpha}\delta_{n\tau}|\mathcal F_{-\alpha}\delta_{n'\tau}>\nonumber\\
&=&<\delta_{n\tau}|\mathcal F_{-\alpha}^\dagger\mathcal F_{-\alpha}\delta_{n'\tau}> \nonumber \\
&=&<\delta_{n\tau}|\delta_{n'\tau}>.
\end{eqnarray}
Therefore, the ChW functions form an orthonormal and complete basis set that continuously depend on $a$, here after called ChW$_a$ basis sets. 

Our approach can be rationalized as a local gauge transformation of the first class~\cite{Pauli:1941}, in which the chirp functions are related to plane-waves  through a gauge transformation of the following form,
\begin{equation}
 e^{iGx} \rightarrow e^{i \phi(x)} e^{i G x} = e^{-i (\eta+k x^2)} e^{i G x}.
\label{equ:gauge_transform}
\end{equation}
Consequently, plane-waves are the special case when the local phase $\phi(x)=0$. This happen when $a=1$. Thus, we can describe the electron wavefunctions in chirp-waves using the following expansion 
\begin{eqnarray} 
\varphi(x)&=&\sum_{n=-\infty}^{\infty} C_{n;\alpha}ChW_{n\tau;\alpha}(x).\label{eq:chirp_serie}
\end{eqnarray}
The expansion coefficients for a given function $\varphi(x)$ can be efficiently obtained through a single discrete fractional Fourier transform ($\mathcal{D}r\mathcal{F}_{\alpha}$) of the sample function $\varphi(x_n)$, as follows
\begin{eqnarray}
C_{n;\alpha}=\frac{e^{i(s(\alpha)\pi/4-\alpha/2)}}{\sqrt{|\sin\alpha|}}f_\alpha(u_n),\label{eq:coeff_serie}
\end{eqnarray}
with $f_\alpha(u_n)=\mathcal{D}r\mathcal{F}_{\alpha}[\psi(x_n)]$. The linear chirps are non-periodic functions, therefore the chirp basis constructed using the above approach can only be used to expand aperiodic functions in the finite interval $[-L/2,L/2]$. However, the electron density is gauge invariant, so it is possible to expand wavefunctions in chirp-waves series. This in general implies
\begin{eqnarray}
\rho(x)=|\varphi_{\alpha}(x)|^2 = |\varphi_{\alpha=\pi/2}(x)|^2,\label{eq:rho}
\end{eqnarray}
which is all what is, in principle, required in density functional theory (DFT) to solve the Kohn-Sham equations~\cite{Kohn:1965}. The theoretical sampling rate to perfectly reconstruct a function in chirp-waves, using its FrFT spectrum, is $\eta=\sin{\alpha/L}$~\cite{Pei:1999}. Nevertheless, in electronic structure calculations a more important quantity is the energy convergence criteria. On the other hand, the contribution of the spectra of a function with compact support become less important as the coefficient index $n$ increases. Therefore, in practice, a perfect reconstruction may not be strictly necessary.


In order to compare the performance of ChW and PW basis sets, we have considered the expansion in series of three $\varphi_{ns}$ orbitals of the Krypton atom. The orbitals were computed within the DFT using the local density approximation (LDA) exchange and the Vosko, Wilk, and Nusair correlation functional (VWN)~\cite{Vosko:1980} using our numerical non-relativistic DFT code for atoms. The radial cutoff distance of $10$~\AA\ was used in all the calculations.

\begin{figure}[htb!] 
\centering
\includegraphics[width=0.8\columnwidth,keepaspectratio=true]{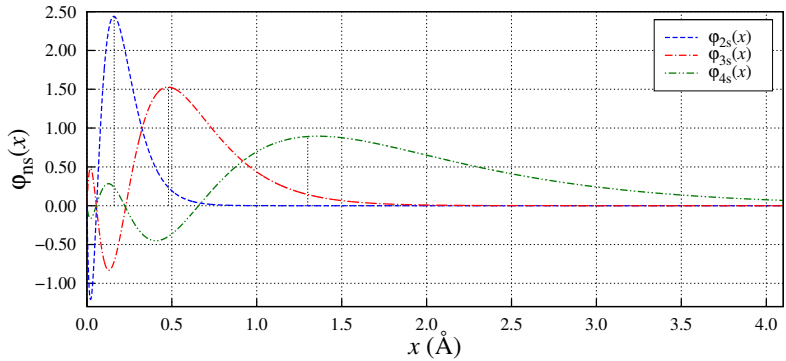}
\caption{\label{fig:kr_orbitals} Radial wavefunctions of the Krypton atom. The vertical dotted lines indicate the core radius.}
\end{figure}

We have evaluated the radial part of $\varphi_{2s}$, $\varphi_{3s}$ and $\varphi_{4s}$ orbitals. This set of orbitals span the conventional partition of the electron states in atoms, i.e., the core, semicore and valence states, and therefore, they will serve to verify the performance of the ChW basis sets for each instance. We computed the fast Fourier transform (FFT) and the discrete FrFT ($a=0.1, 0.2, ..., 0.9$) for each $\varphi_{ns}$ orbital using $900$ equally spaced samples. Only half of the coefficients are necessary because the spectrum is symmetric around the origin for pair functions.

\begin{figure}[t!] 
\centering
\includegraphics[width=0.8\columnwidth,keepaspectratio=true]{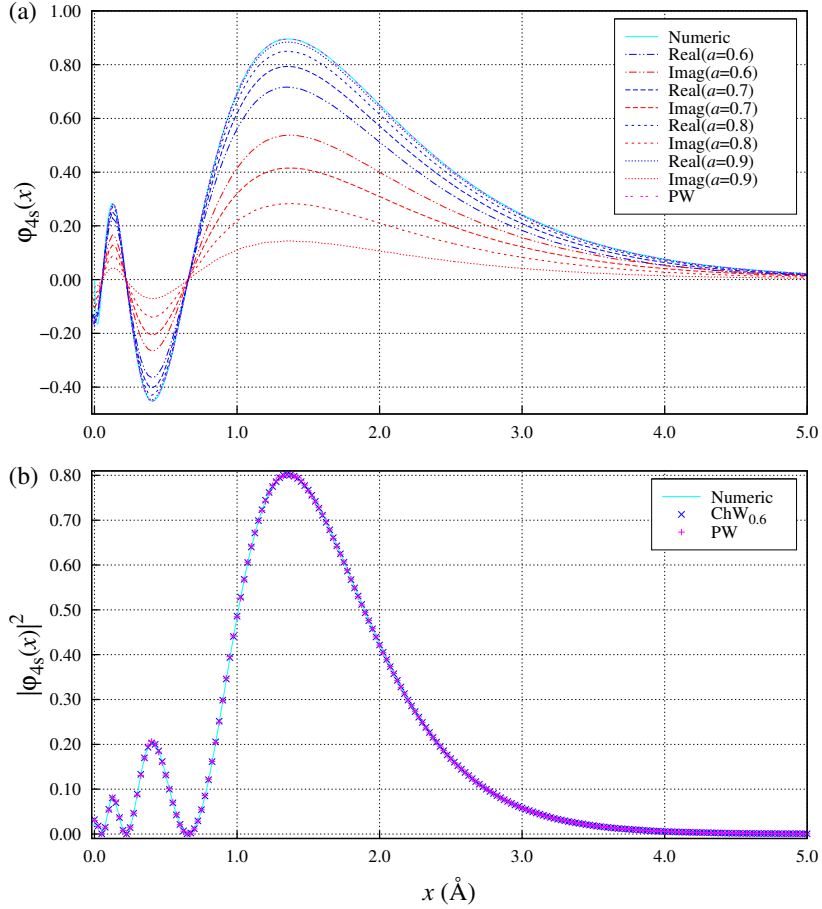}
\caption{\label{fig:phi_rho_alpha} (a) The numeric, ChW$_a$, and PWs represented $\varphi_{4s}$ orbital. (b) The probability density distribution. All the ChW$_{a}$ series overlap, but only the ChW$_{0.6}$ is shown.}
\end{figure}

The chirped characteristics of the functions in the ChW$_a$ basis set is adjusted through the $a$ parameter. As a demonstration, we have reconstructed the $\varphi_{4s}$ orbital using $450$ basis functions of the ChW$_a$ and PW series. The Figure~\ref{fig:phi_rho_alpha} shows the reconstructed $\varphi_{4s}$ orbital for the different series. The sample function and the PWs reconstructed orbital, as expected, are overlapped. On the other hand, the ChW reconstructed functions, according with the definition of chirp series of the equation~(\ref{eq:chirp_serie}), are in general complex, and their real and imaginary components vary with the $a$ parameter. Therefore, the real valued wavefunction might not be reconstructed using ChWs. However, as seen in the Figure~\ref{fig:phi_rho_alpha}(b), because the electron density is gauge invariant, the electron densities are almost identical regardless the value of $a$.

\begin{figure}[t!] 
\centering
\includegraphics[width=0.8\columnwidth,keepaspectratio=true]{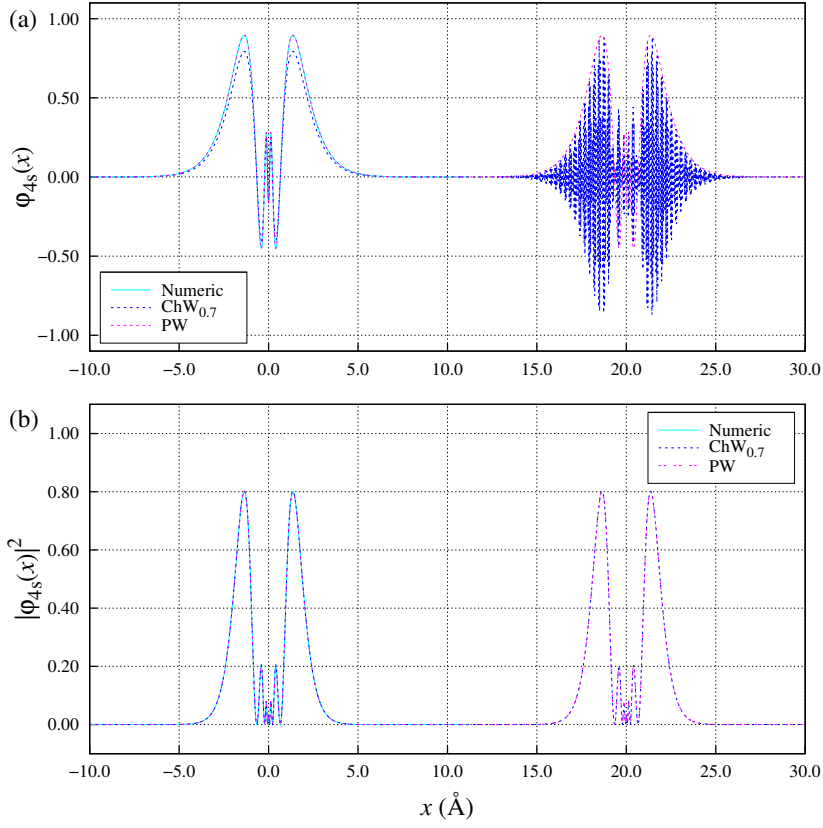}
\caption{\label{fig:phi_gauge} (a) The numeric, real component of the ChW$_{0.7}$ and PW $\varphi_{4s}$ orbital. (b) The electron density in ChWs and PWs are periodic functions.}
\end{figure}

The Figure~\ref{fig:phi_gauge} shows the reconstructed $\varphi_{4s}$ function and its next nearest image on the right side. The plotted real part of the ChW$_{0.7}$ reconstructed orbital is aperiodic, however, the electron density is correctly a periodic function. This result is also independent the of the chosen ChW$_a$ basis set.

\begin{figure}[b!] 
\centering
\includegraphics[width=0.8\columnwidth,keepaspectratio=true]{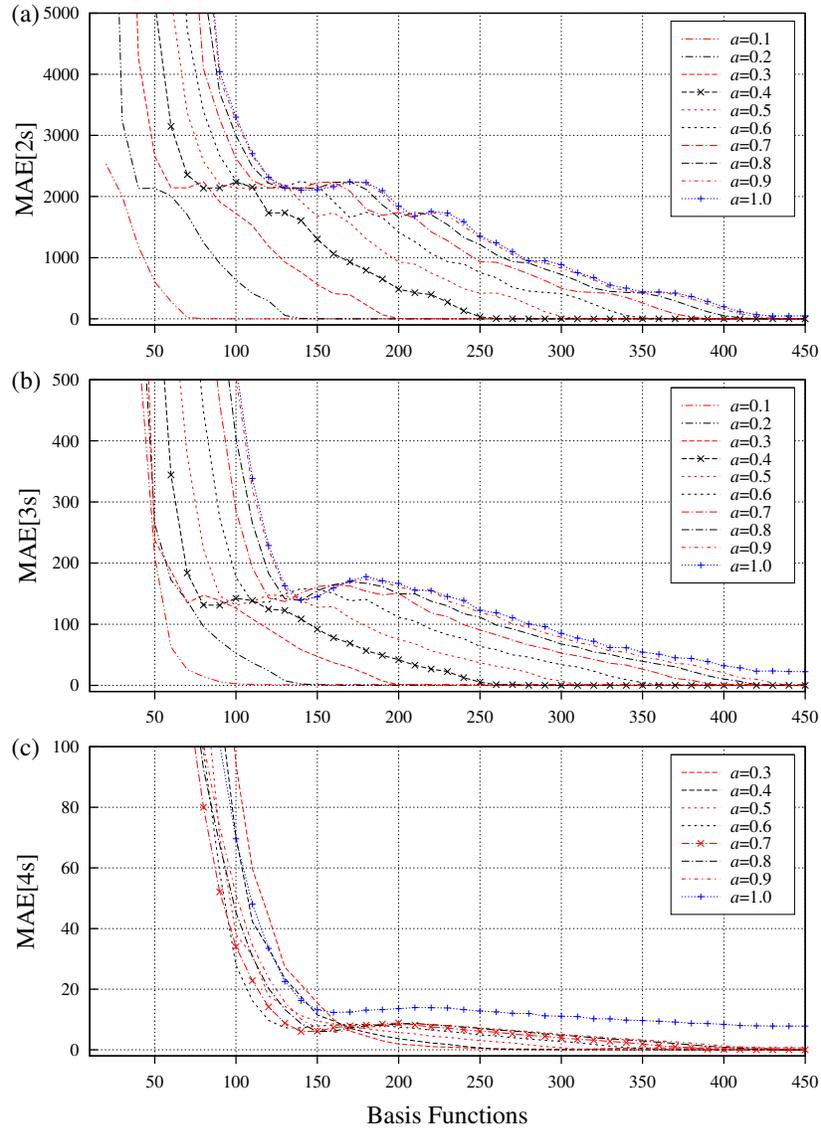}
\caption{\label{fig:mae_core} MAE (in arb. units) of the electron density distribution for the core region. The ChW$_a$ basis set is indicated by $a$, for $a=1$ is the standard PW basis set.}
\end{figure}

In principle, if we know the electron density of a system, the energy can computed within the DFT. For this reason, we have expanded the orbitals using ChW$_a$ and standard PW series employing from $10$ to $450$ basis functions, using an step of $10$. We have evaluated the mean absolute error (MAE), to determine the accuracy in the description of the electron density, using the following expression
\begin{equation}
MAE[ns] = \frac{1}{p}\sum_{i=1}^{p}|\varphi_{ns}^{num}(x_{i})|^2 - |\varphi_{ns}^{\alpha}(x_{i})|^2, \label{eq:mae}
\end{equation}
where $\varphi^{num}$ is the is the numerical computed orbital, and $\varphi^{\alpha}$ is the sample function of the ChW or PW reconstructed orbitals, and $p$ the number of samples. In order to assess the description of the electron density separately. The equation~(\ref{eq:mae}) was independently evaluated for the core and valence regions, using the following core radii, $r_{2s}=0.156$~\AA, $r_{3s}=0.489$~\AA, and $r_{4s}=1.373$~\AA, as illustrated in the Figure~\ref{fig:kr_orbitals}. The evaluated MAEs for the core and valence regions as a function of the number of basis functions are shown in the Figures~\ref{fig:mae_core} and~\ref{fig:mae_valence}, respectively.

\section{Results and Discussion}
\begin{figure}[t!] 
\centering
\includegraphics[width=0.8\columnwidth,keepaspectratio=true]{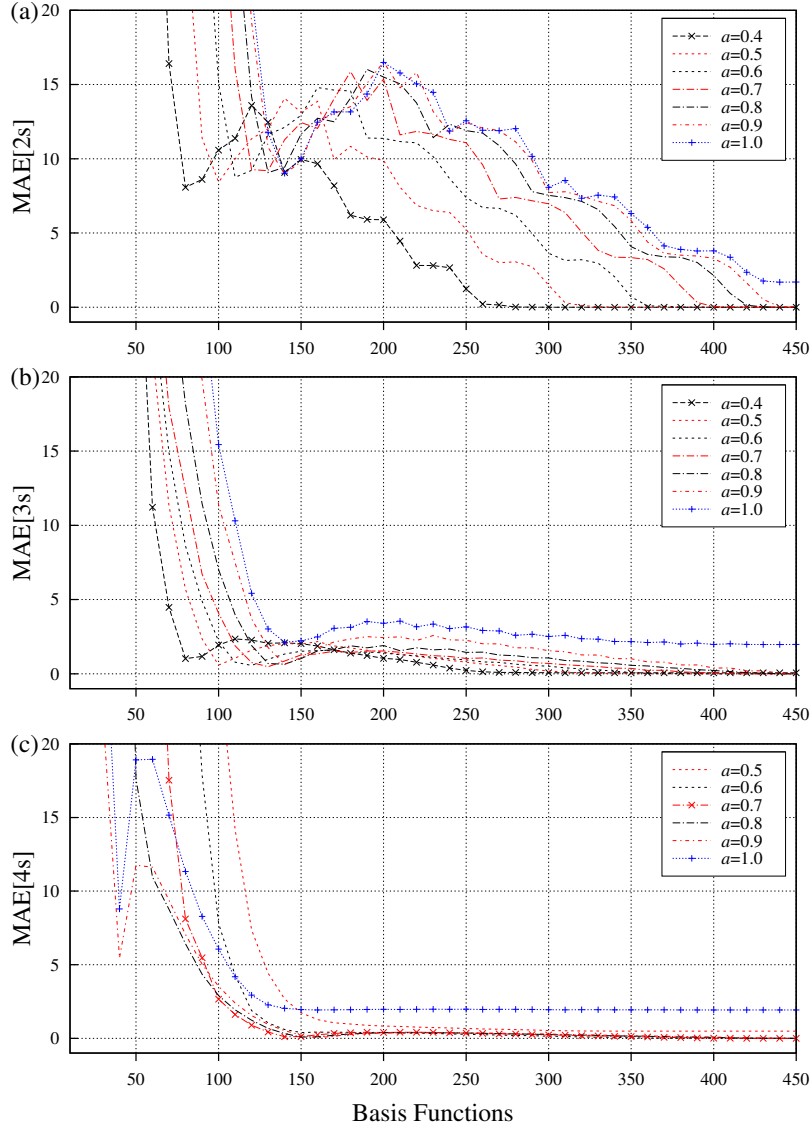}
\caption{\label{fig:mae_valence} MAE (in arb. units) of the electron density distribution for the valence region. The ChW$_a$ basis set is indicated by $a$, for $a=1$ is the standard PW basis set.}
\end{figure}

In general, the MAEs evaluated in the core regions are relatively larger than in the valence regions, up to two order of magnitude. However, from the Figure~\ref{fig:mae_core}, is clear that the convergence of the core \bibliographystyle{abbrv}regions in ChW$_a$ series can be achieved with considerably less terms than with PW series. Indicating that chirp-waves are more suitable functions to describe this region. Furthermore, a very large cutoff may be necessary for the PW series to reach the same convergence as some of the ChWs series. On the other hand, the MAEs for the valence regions are more even, as can be observed in the Figure~\ref{fig:mae_valence}, and only the core orbital $\varphi_{1s}$ require a substantially larger cutoff to converge. Nevertheless, the ChW$_a$ expansions converge prior to the PWs in all the cases. 

The Figures~\ref{fig:mae_core} and~\ref{fig:mae_valence} show that for particular values of $a$, the ChW$_a$ series require a considerable lower cutoff in order to correctly represent the electron density as compared to PWs. Note that this trend becomes more important for core electrons. In fact, the remarkably small cutoff for values of $0.1\leq a<0.4$ suggest a strong chirp-like nature of the core states, indicating that are better approximated by ChWs. In contrast, we found that small values of $a$ were not suitable for the valence region and resulted in very large MAEs, and therefore, were not included in the Figure~\ref{fig:mae_valence}. In order to accurately describe the valence region values of $a\geq 0.4$ were necessary. In this sense, a suitable $a$ may be used in order to correctly describe the core and valence regions simultaneously. In the Figures~\ref{fig:mae_core} and~\ref{fig:mae_valence}, we have highlighted with the symbol $\times$ the curves with our suggested best values of $a$ for each orbital.

The convergence for a particular ChW$_a$ basis set rely on $a$ and the characteristic of the orbital, with more favourable small values of $a$ for the core region than for the valence region. This characteristic makes of the ChWs very attractive, because $a$ can be treated as a variational parameter to search for the optimum ChW$_a$ basis sets that require the lower cutoff.

The chirp-wave series, as defined in the equation~(\ref{eq:chirp_serie}), are not suitable for the representation of the real valued wavefunction due the gauge field, however, the electron density is gauge invariant and can be accurately reconstructed. ChWs share similar advantages as conventional approaches, because chirp-waves span a continuous range in the momentum coordinates, and are not fully localized as plane-waves. In particular, we have found that the natural oscillations of the core states are resembled by linear chirp functions. This explain why chirp-waves are capable of describing the electron density in the core region with a much more efficiency than plane-waves.

\section{Conclusions}
Our results suggest that the chirp-wave basis sets provide particular characteristics that allow for the efficient representation of the electron wavefunctions. In some aspects, chirp-waves may open the bridge for the inclusion of semicore states at low computational cost, eliminating the drawback of approximations. Moreover, the explicit inclusion of core states for accurately relativistic and other core effects, very important in studies with heavy atoms. Allowing the modelling of important properties of atomic systems in which semicore and core states are critically necessary.

As chirp-waves can more efficiently deal with the wavefunctions in the core region, chirp-waves can also take great advantage of pseudopotentials. For instance, ChWs can be used in combination with hard pseudopotentials to extent their transferability to a wide range of chemical environments.

We are currently implementing and testing a DFT/ChW code for atomic clusters in the position coordinates. Future development will be focused in the solution in the FrFT domains. The formulation of the Schr\"odinger equation within the framework of the density functional theory for supercell and band calculations in FrFT position-momentum domains is work in progress. 

\section*{Acknowledgments}
We thanks Prof. Pierre Pellat-Finet. RT thanks to Colciencias for the financial support.


\providecommand{\noopsort}[1]{}\providecommand{\singleletter}[1]{#1}%

\end{document}